\documentclass{article} 
\usepackage{nips15submit_e,times}
\usepackage{hyperref}
\usepackage{url}

\usepackage{amsmath}
\usepackage{amssymb}
\usepackage{graphicx}
\title{Variational auto-encoding of protein sequences}

\author{
Sam Sinai\thanks{Program for Evolutionary Dynamics, Department of Organismic and Evolutionary Biology} \\
Harvard University\\
\texttt{samsinai@g.harvard.edu}
\And
Eric Kelsic\thanks{Wyss Institute} \thanks{To whom correspondence should be directed} \\
Harvard Medical School\\
\texttt{eric\_kelsic@hms.harvard.edu}
\AND
George M. Church\thanks{Department of Genetics} \footnotemark[2]  \footnotemark[3]\\
Harvard Medical School\\
\texttt{church\_lab\_admin@hms.harvard.edu}
\And
Martin A. Nowak\footnotemark[1] \footnotemark[3] \thanks{Department of Mathematics} \\
Harvard University\\
\texttt{martin\_nowak@harvard.edu}
}

\nipsfinalcopy 

\begin{document}

\maketitle

\begin{abstract}

Proteins are responsible for the most diverse set of functions in biology. The ability to extract information from protein sequences and to predict the effects of mutations is extremely valuable in many domains of biology and medicine. However the mapping between protein sequence and function is complex and poorly understood. Here we present an embedding of natural protein sequences using a Variational Auto-Encoder and use it to predict how mutations affect protein function. We use this unsupervised approach to cluster natural variants and learn interactions between sets of positions within a protein. This approach generally performs better than baseline methods that consider no interactions within sequences, and in some cases better than the state-of-the-art approaches that use the inverse-Potts model. This generative model can be used to computationally guide exploration of protein sequence space and to better inform rational and automatic protein design.

\end{abstract}

\section{Introduction}

Protein engineering is of increasing importance in modern therapeutics. Designing novel proteins that perform a particular function is challenging as the number of functional proteins compared to all possible protein sequences is miniscule. This renders naive experimental search for desirable variants intractable. Hence, a computational heuristic that can narrow the experimental search space (virtual screening) is extremely valuable. 

While a variety of energy-based models for protein folding have been used in the past decades, recent advances in machine learning, particularly in the domain of generative models, have opened up new avenues for computational protein design. Rich databases of protein sequences that document functional proteins found in living organisms provide us with ample training data. The majority of these datasets lack labels (indicators of their performance) however, which prompts for an unsupervised learning approach. As these sequences arise from closely related living organisms, it is reasonable to assume that they are functional (and also similar in their functionality).

Given the sparse, unstructured, and discrete space that protein sequences exist in, it is prudent to anchor the search for functional sequences on a known protein with the desired functionality. Starting from that sequence of interest, we can search public databases of sequence variants from related organisms and align them. This alignment of sequences constitutes an evolutionary cluster of nearby variants known as a multiple sequence alignment (MSA). 

We are interested in using MSA data in an unsupervised manner to train models that can inform us about protein function. We hope to then use these models to find good candidate sequences (absent in our training), that can function similarly or better than those that we have already observed. Generative models have two appealing properties for this purpose: (i) They can be trained on sequence data alone, (ii) They can produce new candidate sequences that are similar to those present in our dataset, but not exactly the same. 

Variational auto-encoders (henceforth VAE) \cite{kingma2013,rezende} are one type of such unsupervised generative models that aim to reconstruct their input through a compressed and continuous latent domain. Traditional auto-encoders are neural networks that reconstruct their input imperfectly. VAEs incorporate variational Bayesian methods to impose a lower bound on the probability of an input (which also serves as a built-in regularization method), allowing for a probabilistic interpretation of results. 

A protein sequence $x$ with length $L$ lives on a $L$ dimensional space, each with 20 possible values. The number of sequences within these $20^L$ possibilities that perform a particular function is very small. What we hope to achieve is to compress this large space into a lower dimensional continuous embedding of latent variables that explain the differences between protein sequences. For protein design purposes, we can then traverse this space to find new functional proteins. Additionally, we would hope that this compression would teach us about key properties that affect protein function.

\subsection{Related Work}

Over the past few years generative graphical models have been used on sequence alignments to predict protein structure and function \cite{marks2011protein, ekeberg2013, hopf2017}. These models learn correlations between amino acids in different positions, and then try to approximate the effects of changing one amino-acid to another in a given position. The most successful applications of these methods have used Potts models as their core modeling approach. These models incorporate independent and pairwise interactions along the sequence. The technical details are explained in \cite{marks2011protein, ekeberg2013} and their application for a large set of data has been recently published \cite{hopf2017}. The methods show that harnessing correlations between pairs of amino acids at different positions provides significant power for protein folding and function prediction. Recently, variational-auto encoders have been used for continuous representation of chemical compounds, which allowed for optimization of the process of chemical design \cite{gomez2016automatic}. Additionally, variational inference on graphical models (akin to those presented above as Potts models) were shown to hold promise in predicting protein structure \cite{ingraham2017}. Here we show that VAEs also hold promise for protein design.

\section{Method}

Our model needs to learn the joint probability $p(x,z)=p(z)p(x|z)$ where $z \in Z$ are latent variables, and $x \in X$ are the observed data. If we can learn a good distribution for the latent variables, we can then generate new data points like those in $X$ that we haven't observed in our dataset by sampling $z$ from $p(z)$ and sampling new points $\tilde{x}$ from $p(\tilde{x}|z)$. Computing $p(z|x)$ from our observed data requires us to compute the evidence term for each data point in $X$:

\begin{equation}
p(x)=\int p(x|z)p(z) dz
\end{equation}

A good model would maximize the probability of our data. However, direct computation of this integral is intractable. Instead it can be approximated using variational inference. Specifically, we can approximate $p(x)$ by using the Evidence Lower BOund (ELBO):

\begin{equation}
\log p(x) \geq {\mathbb{E}} _{q}[\log p(x|z)]- D_{KL}(q(z|x)||p(z))
\end{equation}

Where in the above formula, $q$ is a family of normal distributions (a standard assumption of these models), approximating $p(z|x)$, and $D_{KL}$ is the Kullback-Leibler divergence. VAEs learn the parameters $\theta$ and $\phi$ for the distributions $q(z|x)$ and $p(x|z)$ simultaneously through gradient descent. In the  language of neural networks, $q_{\theta}$ specifies the encoder and $p_{\phi}$ specifies the decoder. By maximizing the lower bound on the evidence through gradient ascent, we get an approximation for the maximum likelihood of the data. Notably, we also use the standard assumption that the prior $p(z) \sim {\mathcal{N}}(0,1)$.

Once we have build a generative model that produce sequences like those in our dataset with high probability, we use it to generate novel but similar sequences $\tilde{x}$, or evaluate the likelihood of sequences that the model hasn't seen before.

We treat the one-hot-encoded sequences from the MSA as training data, and train our model to reconstruct these sequences.  Once the model is trained, the probability of each input sequence (from training or test set) can be estimated as follows:

\begin{equation}
\log p(x|z) \propto \log(tr(H^TP))
\end{equation}

Where $H$ is an $m \times n$ matrix representing the one-hot encoding of the sequence of interest, $m$ is the number of amino-acids and $n$ the number of positions considered in the protein alignment. $P$, with identical dimensions as $H$, is the probability weight matrix generated by feeding the network a sequence. $P$ can be generated in multiple ways, but the simplest procedure is to compute it by reconstructing the same sequence that was represented by $H$. Alternatively, $P$ can be computed by an average reconstruction of multiple neighboring sequences, or the reconstruction of the wild-type sequence. We found that these approaches result in similar performance for function prediction.

\subsection{Validation} We validate our model by feeding the network sequences of single and double mutants (with respect to the reference sequence) and calculate their probability. We then compare, through rank correlation, this probability with the experimental fitness measurements. Neither the test sequences, nor the fitness measurements are passed through the network during training. We report the outcomes from training the model using the procedure described above on 5 protein families for which fitness measurements are publicly available \cite{melamed2013, romero2015BG_STR, starita2013UBE, stiffler2015BLAC,rockah2015MTH}.

\subsection{Architecture}

We use the following architecture for the VAE. The encoder and the decoder network have three dense layers of 250 exponential linear units ``ELU'' \cite{clevert2015fast}. The encoder and decoder both include a dropout layer. This architecture was selected by grid search on the hyper-parameters including number of dense hidden layers (1-4), number of units per layer (50-350), inclusion/exclusion of dropout and batch normalization layers between each hidden layer. The final decoder layer uses sigmoid neurons. We use Keras \cite{keras} to implement our VAE model and train our model using the ADAM optimizer \cite{kingma2014adam}. Empirically, networks with dense layers trained faster and performed comparably or better than convolutional layers. For protein function prediction, we use 5 latent variables. For lower dimensional representation and visualization, we use 2 or 3 latent variables. This pruning of latent variables slightly weakens the predictive power (by $0-5\%$ depending on dataset), but provides more easily interpretable representations in the latent space.

\begin{figure}[h]
\begin{center}
\includegraphics[width= 2.55 in]{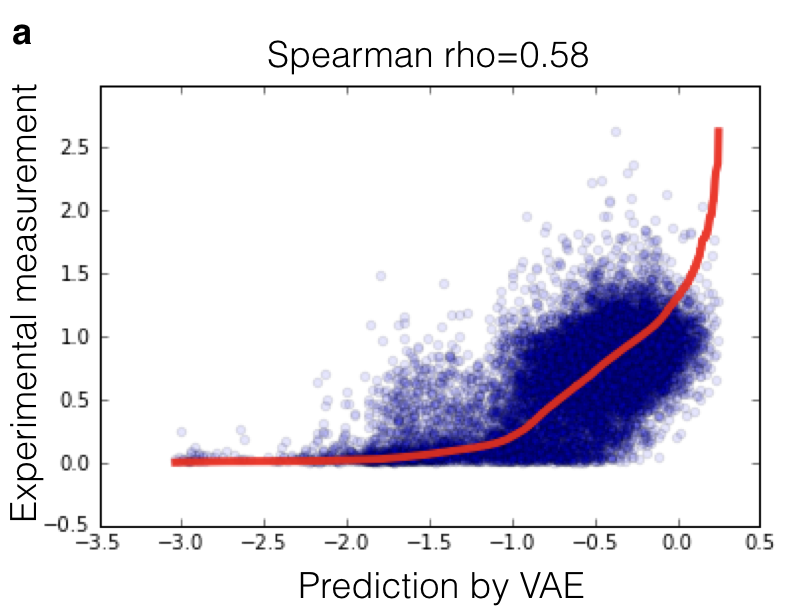}
\includegraphics[width= 2.85 in]{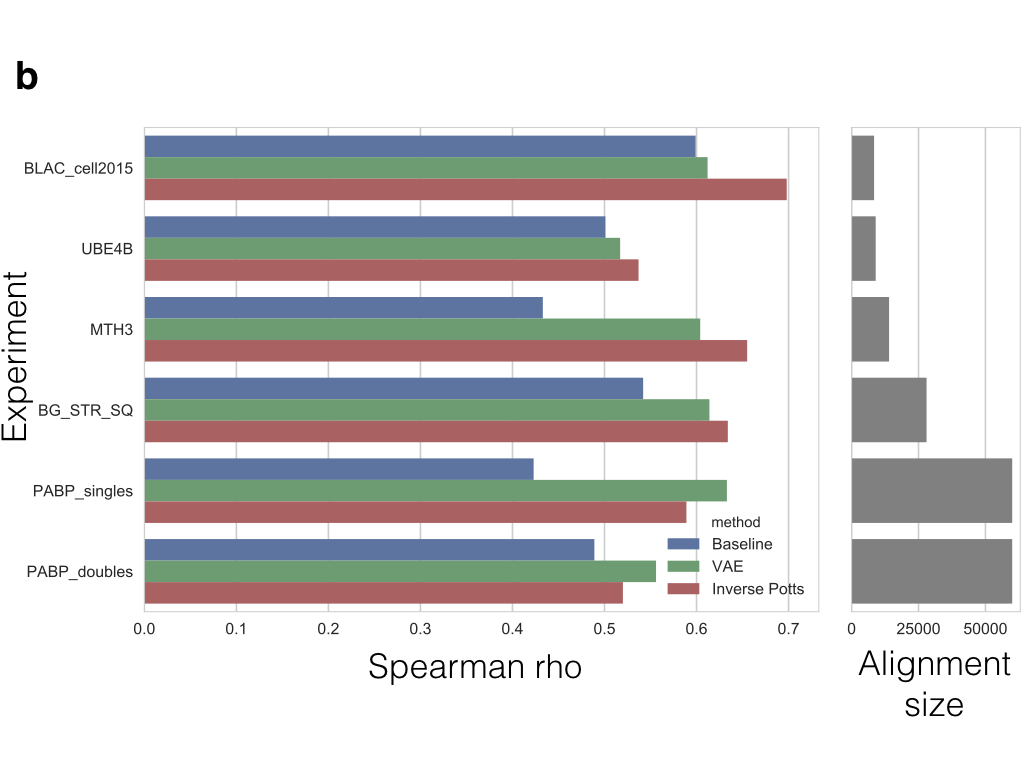}
\includegraphics[width= 2.38 in]{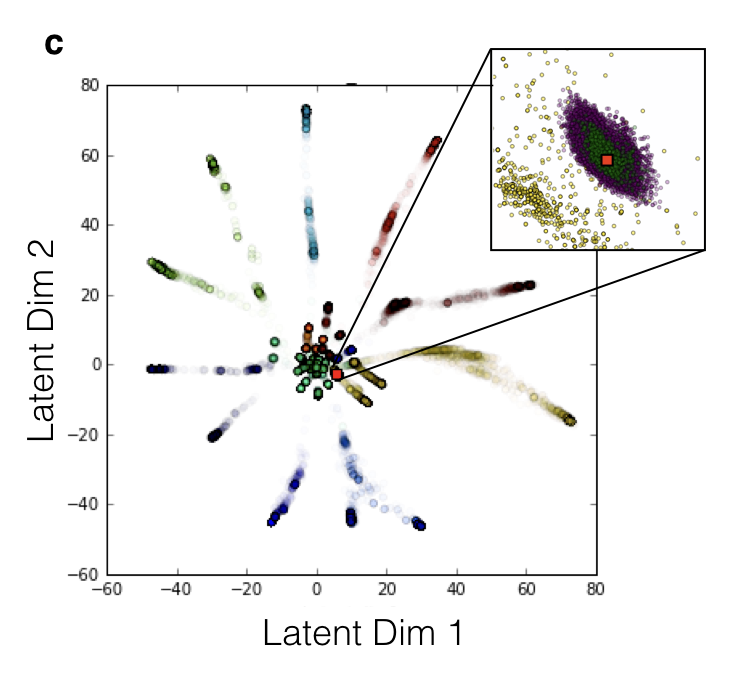}
\includegraphics[width= 3.07 in]{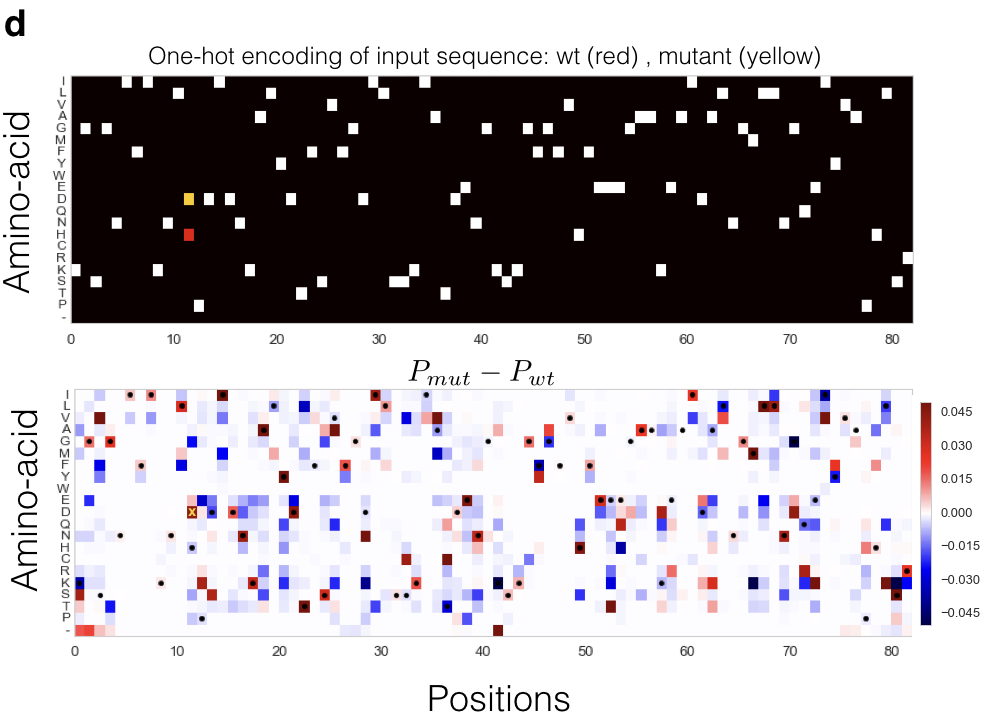}
\end{center}
\caption{\textbf{Summary of results.} (a) Comparison of VAE predictions vs. fitness measurements for double mutations of a reference sequence (wildtype) (Experimental measurements are yeast PABP mutants, see \cite{melamed2013}). Red line shows the hypothetical perfect rank correlation. (b) Comparison of the VAE model's prediction ability with baseline (independent) and pairwise models (Inverse Potts, see \cite{hopf2017}). The size of the dataset is provided for reference. (c) Projection of training data on the 2D latent space. The red square is the latent coordinates for the reference sequence. Points are colored by k-means clustering of sequences, showing that the ``branches" in the star-like latent structure correspond to close-by sequences. This is further confirmed by the fact that the experimental data, single (green) and double (purple) mutants, fall very close to the reference (shown as inset). (d) Top: An example of a change in the input sequence represented as one-hot matrix $H$. Red corresponds to wild-type (reference) and yellow to a mutant (and white is shared positions). These sequences are separately fed into the network, and the difference of the reconstruction matrices $P_{mut}-P_{wt}$ is shown on the bottom panel. Bottom: A single mutation results in updates of probability in many other locations on the sequence, thus at least some pairwise and higher-order interactions (not shown explicitly) are captured. The wild-type sequence is denoted by dark spots and the mutation is marked by x (gold).} 
\end{figure}

\section{Results}

Our results can be summarized by three main observations: (i) The probability estimates calculated by the network  correlate well with protein functions measured in the experiments (Fig. 1a,b)  (ii) The embedding in 2D separates the variants in minimum edit-distance clusters when a 2D latent space is used (Fig. 1c). (iii) The VAE learns pairwise and higher-level interactions between loci on the protein (Fig. 1d) 

We show that VAE's are a viable approach to predict sequence functionality from unlabeled sequences (Fig. 1). Overall, the VAE performs better than the baseline in the 5 datasets tested (Fig. 1b), suggesting that it captures relevant information about protein structure. These datasets were selected because their MSA were presumed to be large and sufficiently diverse for training and because they were used by previous approaches that aimed to predict protein function. We expect that proteins with small MSA size relative to their length and low natural diversity are less suitable for this approach. In line with this expectation, our VAE model performs better than the inverse Potts approach for PABP (for both single and double mutants), which has the largest MSA size relative to its length.  

Our observations indicate that these models can generate candidate sequences that have a high likelihood of performing a particular function comparable to sequences in the training set. Unlike the Inverse Potts model (which it performs closely to), here the latent layer of the VAE provides a continuous representation of the protein sequence. As it has been argued for chemical molecules \cite{gomez2016automatic}, the continuous representation of the protein may be used together with gradient-based optimization to achieve a desirable property. As we show in Fig. 1c, the latent space encodes phylogenetic data (distance clusters), and possibly other features about the protein. The ability to continuously traverse the latent space provided by this approach should yield new opportunities for informed protein design that are qualitatively different than present-day methods.

\section*{Current limitations and future work}

This study serves as a proof-of-concept of the utility of VAEs in representing protein families, as well as their ability to predict the effects of mutations. Our work can be improved in certain dimensions. Despite longer training times, we expect that some recurrent or convolutional architectures may outperform our model, hence a more exhaustive search of such architectures would be prudent. The predicted effects of pairwise and higher order interactions can also be validated by projecting them onto protein tertiary structures. Additionally, our method could be adjusted to use sample weights as is standard in other approaches \cite{marks2011protein, ekeberg2013, hopf2017}. However we found empirically that reweighing did not consistently help the performance across datasets. 

\section*{Acknowledgments}
We would like to thank Pierce Ogden, Surojit Biswas, Gleb Kuznetsov, and Jeffery Gerold for helpful comments. We would also like to thank Debora Marks, John Ingraham, and Adam Riesselman for their feedback on this project as they have independently pursued a similar research objective \cite{riesselman2017deep}. 
\section*{Materials}
Example code to reproduce the analysis from this manuscript on PABP data can be found here: \url {https://github.com/samsinai/VAE_protein_function}

\bibliography{Sinai_Kelsic_Church_Nowak}

\end{document}